\documentstyle[manuscript,prl,aps]{revtex}

\begin{document}
\title{{\bf Ultrafast spin dynamics and critical behavior in half-metallic
ferromagnet : ${\rm Sr_{2}FeMoO_{6}}$}}
\author{T. Kise$^{1}$, T. Ogasawara$^{1}$, M. Ashida$^{2}$, Y. Tomioka$^{3}$, Y.
Tokura$^{1,3}$, and M. Kuwata-Gonokami$^{1,2}$ \cite{auth}}
\address{$^{1}$Department of Applied Physics, the University of Tokyo,Tokyo\\
113-8656,\\
Japan}
\address{$^{2}$Cooperative excitation Project, ERATO, Japan Science and \\
Technology Corporation (JST), Kanagawa 213-0012, Japan}
\address{$^{3}$Joint Research Center for Atom Technology (JRCAT), Tsukuba\\
305-0046,Japan}
\date{\today}
\maketitle

\begin{abstract}
Ultrafast spin dynamics in ferromagnetic half-metallic compound {\rm S}${\rm %
r_{2}FeMoO_{6}}$ is investigated by pump-probe measurements of
magneto-optical Kerr effect. Half-metallic nature of this material
gives rise to anomalous thermal insulation between spins and electrons, and
allows us to pursue the spin dynamics from a few to several hundred
picoseconds after the optical excitation. The optically detected
magnetization dynamics clearly shows the crossover from microscopic
photo-induced demagnetization to macroscopic critical behavior with universal
power law divergence of relaxation time for wide dynamical critical region.
\end{abstract}


\preprint{}

Control and manipulation of spins by ultrafast optical excitation, which
gives rise to photo-induced \ magnetization change and magnetic phase
transitions in dilute magnetic semiconductor quantum structures \cite
{Flytzanis1,Awschalom1,Awschalom2}, doped semiconductors\cite{Awschalom3}
and ferromagnetic metals \cite{Bigot,Hohlfeld,Scholl,Gudde,Beaurepaire1}, 
have attracted considerable attention. Recent study on the 
magnetization dynamics in the photo-excited Ni films with nonlinear optical
techniques has revealed ultrafast spin process within
50 fs \cite{Gudde}. Strongly correlated electron systems with
half-metallic nature, which have perfectly spin polarized conducting
electrons at the ground state\cite{halfmetal}, are promising candidates for
the study of the photo-induced spin dynamics. These materials have been found
to possess exotic physical properties such as colossal magnetoresistance,
which have strong application potential \cite{colossal}. The strong coupling
between spin, charge and lattice degrees of freedom in strongly correlated
systems makes it possible to manipulate the magnetic properties via
cooperative effects induced by optical excitation. In particular, the
evidence of photo-induced phase transition accompanied with magnetization
changes have been recently reported \cite{Miyano,Zhao}. In order to
understand the nature of these phenomena, it is crucial to investigate the
temporal evolution of the spin system in the picosecond time scale. Although
some attempts have been made by employing pump-probe spectroscopy \cite
{Matsuda}, to the best of our knowledge direct investigation of the
ultrafast spin dynamics in half-metallic materials has not been reported so
far. Such an investigation can be carried out by exploiting the time
resolved magneto-optical Kerr effect (MOKE), which has been shown to be a
powerful tool to study the ultrafast dynamics of magnetization \cite
{Bigot,Nurmikko}.

In the present paper we report on the ultrafast pump-probe MOKE and
reflectivity study of dynamics of spin and electron systems in the ordered
double perovskite ${\rm Sr}_{2}{\rm FeMoO}_{6}$. Since Fe$%
^{3+}(3d^{5};t_{2g}^{3}e_{g}^{2},S=5/2)$ and Mo$^{5+}%
(4d^{1};t_{2g}^{1},S=1/2) $ couple antiferromagentically via interatomic
exchange interaction and the down-spin electron of Mo$^{5+}$ is considered
itinerant (upper panel of Fig. 1), a conducting ferrimagnetic ground state
with half-metallic nature is expected for this material. The density
functional calculation\cite{Kobayashi} also shows that the occupied up-spin
band mainly consists of Fe $3d$ electrons, while the Fermi level exists
within the down-spin band composed of Fe $t_{2g}$ and Mo $t_{2g}$ electrons.
The temperature dependence of magnetization and resistivity measurements
under the magnetic field have shown the ferromagnetic phase transition with
the Curie temperature $T_{C}$ $\sim $ 410-450 K\cite{Tomioka} Also, this
material\ in the form of polycrystalline ceramics shows inter-grain
tunneling type giant magnetoresistance at room temperature because of its
spin polarization\cite{Kobayashi}. Another important property of ${\rm Sr}%
_{2}{\rm FeMoO}_{6}${\rm \ }is the enhancement of MOKE due to spin-orbit
coupling of $t_{2g}$\ electrons in the heavy Mo-atom\cite{Shono}. The strong
MOKE signal enables us to investigate the spin dynamics over a wide temporal
range from sub-picosecond to nanosecond and to obtain the critical exponent
of the relaxation time at the magnetic phase transition.

The MOKE measurements are carried out on single crystal ${\rm Sr_{2}FeMoO_{6}%
}$, grown by floating-zone method\cite{Tomioka}, in polar Kerr configuration
under the magnetic field of 2000 Oe, where the magnetization is nearly
saturated at room temperature [15,16], utilizing polarization modulation
by a piezo-elastic modulator (CaF$_{2}$). Figure 1 (a) \ shows the
spectral profiles of ellipticity $\eta $ and rotation angle $\theta $ at
room temperature. A very large MOKE signal, one order of magnitude larger
than that of the doped manganites\cite{Yamaguchi} is observed. The MOKE
signal is proportional to $f\cdot M$, where $M$ is the magnetization and $f$
\ is determined by the complex refractive index at the probe frequency.
Correspondingly, the magneto-optical spectra show resonance, known as the
plasma enhancement effect \cite{Haas}, around 1eV, which is close to the
plasma edge\cite{Tomioka}. The temperature dependence of the $\eta $, probed
at a photon energy of 0.95 eV, clearly shows the magnetic phase transition
at ${\rm 450\ K}$, which is the $T_{C}$ of the present sample\cite{comment}.
Since the reflectivity is almost temperature\ independent in this
temperature range, the sample magnetization can be monitored with the $\eta $.

For the pump-probe measurements, a Ti:Sapphire regenerative amplifier system
(1 kHz repetition rate) with an Optical Parametric Amplifier (OPA) is
used as light source. The second harmonic of the amplified pulses with a
pulse duration of 200 {\rm fs}, photon energy of 3.1 {\rm eV} and a maximum
fluence of 90 ${\rm \mu J/}${\rm cm}$^{{\rm 2}}$ is used as pump
pulse, and its energy is close to the charge transfer excitation from {\rm O}
$2p$ to {\rm Fe/Mo }$t_{2g}$ band with down-spin (see the upper panel of
Fig.1). The probe pulses from the OPA are tuned to 0.95 {\rm eV}, at which
the ellipticity dominates the MOKE signal rather than the rotation effect
(Fig.1 (b)). The pump-probe MOKE measurements are also carried out in polar
Kerr configuration and the polarization change of the reflected light from
the sample is measured by a balanced detection scheme shown in Fig. 2 (a).
By synchronizing the chopper for the pump beam with the regenerative
amplifier, the balanced signal is detected and analyzed shot by shot using
boxcar integrator and A/D converter. The photo-induced Kerr
ellipticity change is measured as the difference between magnetization
reversal signals, i.e., $\Delta \eta _{Kerr}=\frac{1}{2}[\Delta \eta
(M)-\Delta \eta (-M)]$ by changing the sign of the magnetic field in order
to eliminate the contribution from the pump induced optical anisotropy. A
sensitivity of 10$^{-3}$\ deg is achieved in our measurement system. The
signal is observed to be proportional to the pump beam intensity in all
pump-probe measurements.

The inset in Fig. 2 (a) shows the transient reflection change $\Delta R/R$,
measured at 300 {\rm K}. It shows a sharp reduction in the reflectivity
during the pump pulse duration, followed by a fast relaxation within ${\rm %
2\sim 3 ps}$ (region (1)) and a fairly long time plateau up to few tens of
nanosecond (region (2)). The reflectivity returns to the initial state in 
{\rm 1 ms} by heat diffusion (region (3)). \ Figures 2 (b) and (c) show the $%
\Delta R/R$ and ellipticity change $\Delta \eta _{Kerr}$ for different
temperatures, indicating that the temperature dependence is negligible\ in $%
\Delta R/R$, while that is significant in $\Delta \eta _{Kerr}$. The temporal
evolution of $\Delta \eta _{Kerr}$ up to 500 ps is shown in Fig.3 (a) for
different temperatures. One can observe from Fig. 3(a), that below the Curie
point, it can be fitted by $\Delta \eta _{Kerr}\left( t\right) =\Delta \eta
_{step}+(\Delta \eta _{max}-\Delta \eta _{step})(1-\exp \left[ -t/\tau
_{spin}\right] )$, where $\Delta \eta _{step}$ describes the instantaneous
decrease in the $\Delta \eta _{Kerr}$, while $\Delta \eta _{max}$ is the
asymptotic value at the quasi-equilibrium state. The signal, $-\Delta \eta
_{max}$, increases drastically close to $T_{C}$ as shown in Fig.3 (b). Our
measurements reveal nearly linear increase of $-\Delta \eta _{max}$\ with
the pump intensity.

The most striking feature is the very slow spin thermalization observed in $%
\Delta \eta _{Kerr}$ signal in comparison with the electron thermalization
observed in transient $\Delta R/R$ data. In ${\rm Sr}_{2}{\rm FeMoO}_{6}$,
the behavior of electrons is similar to that of ferromagnetic nickel \cite
{Bigot}. Specifically, the electron temperature rises rapidly by the optical
excitation and it relaxes within 2 to 3 picoseconds to reach
quasi-equilibrium temperature, which is 8-10 K, obtained from $\Delta R/R$,
higher than the initial temperature. The fast decay of transient
reflectivity indicates that the local heat transfer from electron to the
lattice system is completed within a few picoseconds, accompanied by the
lattice heat-up to reach quasi-equilibrium temperature. On the other hand,
the behavior of the spin system in ${\rm Sr}_{2}{\rm FeMoO}_{6}$ looks very
different from the behavior of the electronic system. Specifically, the very
slow spin thermalization (see Fig. 2(c)), which is pronounced at higher
temperature, indicates the anomalously small heat exchange between electrons
and spins in ${\rm Sr}_{2}{\rm FeMoO}_{6}$. Such an electron-spin thermal
insulation can be attributed to the nature of the half-metallic electronic
structure where conducting electrons are perfectly spin polarized in the
down-spin band and isolated from the insulating up-spin band as shown
schematically in the upper panel of Fig. 1. Thermal motion of electrons
around the Fermi level in the spin polarized conduction band does not
increase the spin temperature.

From the observed results, we have the following scenario for the temporal
evolution of electron, lattice and spin system in ${\rm Sr}_{2}{\rm FeMoO}_{{%
6}}$. Initially, during the photo-excitation (${\rm \leq 1\ ps}$), the
electron system is heated-up and rapidly thermalized due to
electron-electron interaction. In this first stage the ellipticity shows a
sharp decrease ($\Delta \eta _{step}$). In the next stage, the electron
system relaxes by its energy transfer to the lattice system. The electron
and lattice systems reach quasi-equilibrium state (${\rm \sim 5 ps}$) by the
electron-phonon interaction, leaving the spin system at its initial
temperature. After that, the spin slowly relaxes toward this
quasi-equilibrium state through weak heat exchange with the reservoir at
quasi-equilibrium temperature. Finally, the system returns to the initial
state by heat diffusion.

The sharp decrease in the ellipticity is the major feature of the
initial stage of the optical relaxation in ${\rm Sr}_{2}{\rm FeMoO}_{6}$%
\ (see inset of Fig. 2). The ratio $\Delta \eta _{step}/\eta $ shows a 
weak temperature dependence (see Fig. 3(c)). Since MOKE signal
is proportional to $f\cdot M$, both photo-induced change in the
refractive index ($\Delta f/f\sim \Delta R/R)$ \ as well as the
photo-induced magnetization change $\left( \Delta M/M\right) $ \
contribute to $\Delta \eta _{step}/\eta $. Though the relative
instantaneous changes in the ellipticity and reflectivity are of the same
order, $\Delta \eta _{step}/\eta \sim \Delta R/R\sim 0.01$, the
subsequent temporal evolution of $\Delta \eta _{Kerr}$ in the
picosecond time scale is very different from that of $\Delta R$. This
indicates the direct demagnetization by resonant optical excitation. However,
as it has been discussed in recent papers on Ni \cite{Bigot,Hohlfeld,Scholl,Gudde}, it is premature to directly connect the MOKE signal with 
demagnetization in such ultrafast time scale.

We now discuss the temporal evolution of the MOKE signal in the second
stage, when the electron and lattice system have reached the
quasi-equilibrium (plateau region (2) in the inset of Fig. 2). The dramatic
increase in $\Delta \eta _{max}$(see Fig. 3(b)) and the relaxation time $%
\tau _{spin}$ as the temperature approaches $T_{C}$ indicate that the time
resolved signal directly reflects the critical behavior of magnetization at
the ferromagnetic phase transition. The spin temperature at the
quasi-equilibrium, which can be estimated from Fig. 3(b) and the temperature
dependence of the $\eta $ (Fig. 1(a)), is in good agreement with the
electron temperature estimated from $\Delta R/R$.

It is necessary to emphasize that the time-resolved MOKE measurements give
us an unique opportunity to study the critical dynamics of spin system
independently from other degrees of freedom and obtain the critical
characteristics of the ferromagnetic phase transition. The dynamics of the
second order phase transition can be described by the dynamical scaling
theory \cite{Fisher,Suzuki,Miyashita}, which allows us to relate the
critical behavior of the kinetic parameters (e.g. the relaxation time) to
the critical exponents of the static parameters (e.g. correlation length) on
the both sides of the critical point. The theory predicts that in the
vicinity of $T_{C}$, the relaxation time of the order parameter can be
described as $\tau \propto |T-T_{C}|^{-z\nu }$, where $\nu $$\ $and $z$
denote the critical exponent of the correlation length and the dynamical
critical exponent respectively. Figure 4 shows the temperature dependence of 
$\tau _{spin}$ as a function of $|T^{\prime }-T_{C}|$, where $T^{\prime }$%
denotes the quasi-equilibrium temperature at the plateau region (temperature
region (2) in the inset of Fig.2(a)). One can clearly observe the power law
divergence with $z\nu =1.22\pm 0.06$ for the spin relaxation time in the
vicinity of $T_{C}$\cite{critical}. It should be emphasized that the power
law behavior is established in the time scale of few tens of picosecond,
while the width of the dynamical critical region is much higher than for
conventional metals \cite{Huang}. The theoretical calculation for the three
dimensional Ising and Heisenberg models give $z\nu \approx 1.30$\cite{Ito}
and $1.37$\cite{Pesczak}, respectively, while the two-dimensional Ising
model gives $z\nu \approx 2.165$\cite{twodising}. Therefore, our
measurements clearly indicate three dimensionality of the spin system in $%
{\rm Sr_{2}FeMoO_{6}}$.

We have presented ultrafast spin dynamics in the ordered double perovskite $%
{\rm Sr_{2}FeMoO_{6}}$ by using the time-resolved MOKE technique. We have
observed, for the first time, extremely slow relaxation of spins, thermally
insulated from electron and lattice systems due to the half-metal nature of
this material. The thermal insulation of spin system provides us a unique
opportunity to examine the non-equilibrium spin dynamics near the critical
point in a time scale from picosecond to nanosecond range. Crossover from
ultrafast microscopic spin relaxation to macroscopic critical behavior has
been clearly demonstrated. In the vicinity of the critical point, the spin
relaxation time increases as $|T-T_{C}|^{-(1.22\pm 0.06)}$ , which is
consistent with the theoretical prediction for the 3D ferromagnetic system. 
We also observe very fast decrease in the MOKE signal $\Delta \eta
_{Kerr}$ caused by charge transfer optical excitation. Although the
underlying physical mechanism of such an ultrafast phenomenon is not
established yet, the distinct difference in the temporal profiles of the
photo-induced reflectivity and ellipticity suggests the contribution from
ultrafast spin dynamics to nonlinear MOKE signal. 

The authors are grateful to S. Miyashita, N. Ito, N. Nagaosa, M. Ueda, Yu.
P. Svirko and C. Ramkumar for illuminating discussions. This work is
supported in part by a grant-in-aid for COE Research from the Ministry of
Education, Science, Sports and Culture of Japan and the New Energy and
Industrial Technology Development Organization (NEDO).

\clearpage

Figure captions.

\bigskip

Fig. 1. Magneto-optical Kerr measurements on ${\rm Sr_{2}FeMoO_{6}}$ \ under
the magnetic field of 2000 Oe.(a) Kerr rotation (solid line) and ellipticity
(dashed line) spectra at 300K. (b) Temperature profile of linear Kerr
ellipticity probed at 0.95eV. The upper panel shows the spin configuration
of Fe and Mo ions and a schematic of the electronic band structure of ${\rm %
Sr_{2}FeMoO_{6}}$ based on the density-functional calculations by Sawada and
Terakura (Ref.\cite{Kobayashi}).

\bigskip

Fig. 2. (a) Schematic of the experimental setup for the pump-probe
magneto-optical Kerr measurements. Inset shows the temporal evolution of
transient reflection change $\Delta R/R$ from subpicoseconds up to
millisecond. Temporal evolution of the transient reflection $\Delta R/R$ (b)
and Kerr ellipticity change $\Delta \eta _{Kerr}$ (c) up to 50ps, measured
at 200K, 300K and 400K.

\bigskip

Fig. 3. (a) Temporal evolution of photo-induced Kerr ellipticity $\Delta
\eta _{Kerr}$ up to 500 ps, measured at various temperatures. Solid lines
are the exponential fit. Temperature profiles of $\Delta \eta _{max}$ (b)
and rapid component $\Delta \eta _{step}$ normalized to linear Kerr
ellipticity $\eta $ (c).

\bigskip

Fig. 4. Temperature dependence of spin relaxation time $\tau _{spin}$ as a
function of $|T^{\prime }-T_{C}|$. The solid line is a power law fit $%
|T^{\prime }/T_{C}-1|^{-z\nu }$ for the points near $T_{C}.$ The fit returns 
$z\nu =1.22\pm 0.06$.

\clearpage
\begin{figure}[tbp]
\end{figure}

\clearpage
\begin{figure}[tbp]
\end{figure}

\clearpage
\begin{figure}[tbp]
\end{figure}

\clearpage
\begin{figure}[tbp]
\end{figure}


\begin{references}
\bibitem[*]{auth}  Author to whom correspondence should be addressed.
Electronic address: gonokami@ap.t.u-tokyo.ac.jp

\bibitem{Flytzanis1}  C. Buss {\it et al.,} Phys. Rev. Lett. {\bf 78}, 4123
(1997); M. Haddad {\it et al.,} Appl. Phys. Lett. {\bf 73}, 1940 (1998).

\bibitem{Awschalom1}  S. A. Crooker {\it et al.,} Phys. Rev. B {\bf 56},
7574 (1997).

\bibitem{Awschalom2}  J. J. Baumberg {\it et al.,} Phys. Rev. B {\bf 50},
7689 (1994).

\bibitem{Awschalom3}  J. M. Kikkawa and D. D. Awschalom, Phys. Rev. Lett. 
{\bf 80}, 4313 (1998); Nature {\bf 397}, 139 (1999); Science {\bf 287}, 473
(2000).

\bibitem{Vaterlaus}  A. Vaterlaus, T. Beutler, and F. Meier, Phys. Rev.
Lett. {\bf 67}, 3314 (1991).

\bibitem{Bigot}  E. Beaurepaire, J. -C. Merle, A. Daunois, and J. -Y. Bigot,
Phys. Rev. Lett. {\bf 76}, 4250 (1996).

\bibitem{Hohlfeld}  J. Hohlfeld, E. Matthias, R. Knorren, and K. H.
Bennemann, Phys. Rev. Lett. {\bf 78}, 4861 (1997).

\bibitem{Scholl}  A. Scholl, L. Baumgarten, R. Jacquemin, and W.Eberhardt,
Phys. Rev. Lett. {\bf 79}, 5146 (1997) .

\bibitem{Gudde}  J. Gudde, U. Conrad, V. Jahnke, J. Hohlfeld and E.
Matthias. Phys. Rev. B {\bf 59}, 6608 (1999).

\bibitem{Beaurepaire1}  E. Beaurepaire, M. Maret, V. Halte, J. -C. Merle, A.
Daunois, and J. -Y. Bigot, Phys. Rev. B {\bf 58}, 12134 (1998).

\bibitem{halfmetal}  J. -H. Park {\it et al.,} Nature {\bf 392}, 794 (1998).

\bibitem{colossal}  For a review, Colossal Magnetoresistive Oxides, Ed. by
Y. Tokura, (Gordon \&Breach Publishers, 2000).

\bibitem{Miyano}  K. Miyano, T. Tanaka, Y. Tomioka, and Y. Tokura, Phys.
Rev. Lett. {\bf 78}, 4257 (1997).

\bibitem{Zhao}  Y. G. Zhao {\it et al.,} Phys. Rev. Lett. {\bf 81}, 1310
(1998).

\bibitem{Matsuda}  K. Matsuda, A. Machida, Y. Moritomo, and A. Nakamura,
Phys. Rev. B {\bf 58}, R4203 (1998).

\bibitem{Nurmikko}  Ganping Ju {\it et al}., Phys. Rev. Lett. {\bf 82}, 3705
(1999).

\bibitem{Kobayashi}  K. -I. Kobayashi {\it et al.,} Nature {\bf 395}, 677
(1998).

\bibitem{Tomioka}  Y. Tomioka {\it et al.,} Phys. Rev. B {\bf 61}, 422
(2000).

\bibitem{Shono}  K. Shono, M. Abe, M. Gomi and S. Nomura, Jpn. J. Appl.
Phys. {\bf 20}, L426 (1981).

\bibitem{Yamaguchi}  S. Yamaguchi, Y. Okimoto, K. Ishibashi, and Y. Tokura,
Phys.Rev. B {\bf 58}, 6862 (1998).

\bibitem{Haas}  H. Feil and C. Haas, Phys. Rev. Lett. {\bf 58}, 65 (1987).

\bibitem{comment}  The slight deviation of $T_{C}$ from the results of ref. 
\cite{Tomioka} may be due to the improved site order of Fe and Mo, which is
sensitive to the sample preparation and annealing temperature of ref. \cite
{Tomioka}.

\bibitem{Fisher}  M. E. Fisher and M. N. Barber, Phys. Rev. Lett. {\bf 28}%
,1516 (1972); M. N. Barber, in {\it Phase Transition and Critical Phenomena}%
, vol. 8 ed. C. Domb and J. L. Lebowitz, (Academic Press, London 1983).

\bibitem{Suzuki}  M. Suzuki, Phys. Lett. A {\bf 58}, 435 (1976); Prog.
Theor. Phys. {\bf 58, }1142 (1977){\bf .}

\bibitem{Miyashita}  S. Miyashita and H. Takano, Prog. Theor. Phys. {\bf 73}%
, 1122 (1985).

\bibitem{critical}  In the close vicinity of the critical point the system
is very sensitive to external fields and the possible thermal drift effects
would be prononced. Correspondingly, in order to minimize the possible
uncertainty in the estimation of the ctitical exponent, we intentionally
avoid the experimental point closest to $T_{C}$ in Fig. 4 for fitting.

\bibitem{Huang}  K. Huang, in {\it Statistical Mechanics,} 2nd ed. (John
Wiley \& Sons, New York 1987) p.437.

\bibitem{Ito}  N. Ito, Physica A {\bf 192}, 604 (1993).

\bibitem{Pesczak}  P. Peszak and D. P. Landau. J. Appl. Phys. {\bf 67}, 5427
(1990).

\bibitem{twodising}  N. Ito, Physica A {\bf 194}, 591 (1993).
\end{references}
\end{document}